\documentstyle[12pt]{article}
\include{epsf}
\begin{document}

\title{\Large \bf Aging process of electrical contacts \\in granular matter}

\author{ \large \bf S.Dorbolo,$^1$ M.Ausloos,$^1$ N.Vandewalle,$^1$ M. Houssa$^2$
\\  \\$^1$ GRASP, Institut de Physique B5, Universit\'e de Li\`ege, \\
B-4000 Li\`ege, Belgium 
\\ $^2$ L2MP, Bat. IRPHE \\ 49, rue Joliot Curie \\ BP 146
Technop\^ole de Ch\^ateau - Gombert \\
13384 Marseille Cedex 13, France}

 \maketitle

\begin{abstract}

The electrical resistance decay
of a metallic granular packing has been measured as a function of
time.  This measurement gives information about the size of the
conducting cluster formed by the well connected grains.  Several
regimes have been encountered.  Chronologically, the first one
concerns the growth of the conducting cluster and is identified to
belong to diffusion processes through a stretched exponential
behavior.  The relaxation time is found to be simply related to the
initial injected power.  This regime is followed by a reorganisation
process due to thermal dilatation.  For the long term behavior of the
decay, an aging process occurs and enhances the electrical contacts
between grains through microsoldering.

\end{abstract}

\vskip 2cm

\newpage 

  \section{Introduction}

  The structure of granular materials is characterized by arches that
distribute the weight of the packing towards the edges of the vessel.
This induces an anisotropic distribution of pressures at the points
of contacts between grains.  The electrical resistance $R$ of one
contact has been found to depend on the applied pressure, as a power
law as demonstrated by Hertz \cite{hertz}.  Any variations of the
structure of the heap, whatever its nature (geometrical, chemical,
electrical and mechanical), changes the contact network and in
particular the overall electrical resistance, indeed in a way very
sensitive to external perturbations \cite{branly,onesti}.  The
information contained in $R$ is also linked to the nature of the
contacts.  The surface of metallic grains may often be covered by an
oxide which introduces some semiconducting properties to the
contacts.  This chemical layer in fact constitutes a tunneling
barrier to the current flow.  Therefore a granular packing may be
represented by a skeleton of electrical links between grains, each
link being characterized by a certain electrical resistance
determined by the pressure between beads (given by Hertz relation)
\cite{kirk,takayasu,roux}. The electrical resistance can thus be a
tool which allows to explore arching pressure distribution without
markedly changing the structure.   Such a tool can be useful in
probing stress in granular system as they are submitted to a shear
for example.  Moreover the study of the electrical properties of
granular material can give information concerning large electrical
complex network.

  Recent studies have shown that the electrical properties of metallic
granular packings are complex, e.g. the voltage $V$ across a sample
is not univocally defined by the injected current $i$.  Indeed,
current-voltage $i-V$ curves are non-linear, sometimes singular or
even discontinuous and exhibit memory effects
\cite{roux1,bonamy,apl}.  A typical behavior is shown in Fig. 1.  The
role of the grain oxide layer has been pointed out to be responsible
for the non-linear behavior of the electrical resistance with respect
to the injected current \cite{apl}.  Moreover taking into account the
structure of the network and the nature of the contacts, the shape of
the $i-V$ curves  and its discontinuities (Fig.1) may be explained
\cite{apl}.  The sharp transitions, first discovered by
Calzecchi-Onesti \cite{onesti}, occur when the current reaches a
certain value at which the resistance falls several orders of
magnitude.  This process is very similar to breakdowns in
semiconducting devices \cite{mich}.  This dielectric breakdown
implies high local electronic reactivity, increase in temperature and
grain soldering.  However, microsolderings between grains may occur
before a Calzecchi-Onesti transition \cite{futur}, just like
nucleation sites are initiated by high concentration gradients at a
first order phase transition.  Those solderings are non equilibrium,
or irreversible processes thereby being the fundamental cause for the
hysteretic behavior found in $i-V$ curves of granular packing.

We have recorded the natural evolution of the electrical resistance
of a lead beads heap versus time when a stable current flows through
the system.  A fixed current is injected during 20 min and the
voltage is sampled every 0.2 s.  Different current intensities have
been injected in order to compare their effect on the variations of
the resistance with time.  We emphasize that this situation differs
drastically from a classical $i-V$ curve measurement since in this
latter case, the steps for increasing the injected current are very
small and their succession very slow.  The voltage has been found to
decrease by about 30\% depending on the value of the injected current.

The experimental set up is described in Sect.2.  A discussion is
found in sect.3.  The conclusions are drawn in Sect.4.

\section{Experimental set up}

The experimental set up is simple but the different operations have
to be performed with care because of the high sensitivity of the
system to external perturbations.   The lead beads have a mean
diameter of $2.35$ mm with a polydispersity of $2\%$.  Energy
Dispersive X-ray measurements have shown that the surface is made of
a nanometric layer of lead oxide.  About 14,000 beads have been
stacked into a $50 \times 50 \times 40$ mm$^3$ plexiglas
parallelepipedic vessel.  The density of such a 3D packing is about
0.75 and is naturally near the close packing configuration.
Electrical current leads are two lead rectangular plates placed on two
opposite faces of the vessel and connected to a Keithley K2400
current source \cite{apl}.  The voltage $V$ has been recorded at a 5
Hz frequency on a computer.  The whole system was placed in a
climatized Faraday cage at 16¡C; no vibration was expected to occur
during the experiments.

\section{Results and interpretations}

One of the typical $i-V$ behavior of the lead bead heap has been
reported in Fig.1.  Different features which have been discussed
above are indicated in the figure, i.e. (i) hysteretic loops due to
microsoldering, (ii) Calzecchi-Onesti transition and (iii) a peak
indicating the sensitivity of the electrical resistance to mechanical
shocks.

Figure 2 presents $R(t)$ results obtained for different injected
currents between 10 and 65 mA.  The resistance $R$ has been obtained
by dividing the measured voltage $V$ by the injected current $i$.
The different lines smoothly link the experimental data points.
Symbols are placed every 50 points for trend emphasis.

The resistance decreases with time whatever the injected current is.
This strongly resembles the behavior obtained in the case of the
degradation of ultrathin oxide layers in presence of a electrical
stress.  Indeed such a behavior of the voltage can be found in ref.
\cite{mich,pennetta}.  Nanolayers of insulator stressed by a voltage
are studied in those works.  Leakage currents induce damage to the
insulator that allows more and more current to pass through.  In the
present case, the thin oxide layer that covers the beads is
responsible for such a behavior.

The initial resistance differs according to the injected current.
The higher the injected current, the lower the electrical resistance.
This matches with the non-linear behavior observed in the $i-V$
curves in Fig.1 where the slope is proportional to the differential
resistance.  During the first 2000 seconds of observation the
resistance keeps on decreasing towards some saturating value.

The evolution of the resistance for an injected current of 65 mA is
particular : a Calzecchi-Onesti transition occurs after 250 s, i.e.
the resistance falls by one order of magnitude from 1800 $\Omega$ to
35 $\Omega$.  In this case, two processes are noticed : (i) a slow
process which governs the decay of the resistance with time and (ii)
an avalanche process which induces the Calzecchi-Onesti transition
(like a Zener breakdown).

In Fig.3, the voltage $V$ normalized to the initial voltage
$V_0=V(t=0)$ has been represented versus time in a semi-log plot for
four different currents ($i < 50$ mA), as indicated by the legend (a
normalized voltage is equivalent to a normalized resistance). The
decay amplitude is larger when the current is low, e.g. for $i=10$
mA, the decay is about 25 $\%$.  Chronologically, the voltage decay
may be decomposed into three phases.  (i) During the first minute,
the slope of the R-decay is about the same whatever the current value
is.    (ii) The curves obtained for $i=$ 10 and 25 mA seem to
regularly decrease.  The curves obtained for $i=$ 35 and 50 mA
stabilise around $V/V_0 =$0.82 during 1000 s before decreasing again.
(iii) The slope of the curves are found to be the same from the
1000th second to the end of the experiment.  We may conclude that
various processes occur.

To sum up, the decay process goes as follows : (i) a primary voltage
decrease during the first minutes, (ii) a temporisation stage (due to
the competition of two kinds of effects), especially seen in curves
$i=$ 35 and 50 mA and (iii) a secondary decay phase.

In Fig.4, the data are analyzed during the first phase.  A stretched
exponential fit,
\begin{equation}
\frac{V}{V_0}=(1-a)+a \/ e^{- \sqrt{t/\tau}}
\end{equation} is found where $a$ and $\tau$ are fitting parameters
(see inset).  The square root exponent characterizes a diffusion
process as in crystal growth process \cite{avrami,gadomski,avramie}.
In the present case, this exponent can be interpreted as follows :
The conduction of the current through the heap is thought to be done
through a cluster of `well' connected beads.  If such privileged
contacts exist, they shunt the remain of the packing.  Recall that
electrical stresses are concentrated along this cluster path.  This
means that any enhancement of other contacts allows the connected
cluster to grow as in an avalanche process.  This process is related
to a diffusion of stresses through the sample.  Pennetta et al. show
that such an avalanche-like propagation exists in semiconductor
integreation technology.  Some percolation models have been
developped.  However, the resistance decrease seems to be smoother
than the one found in percolation models \cite{penneta}.

The cluster growth is limited by the size of the vessel; this can
explain a saturation process.  This first stage is shortened as the
current is increased.  The caracteristic time $\tau$ depends on the
value of the initial injected power $P=V_0 i$ as $\tau \propto
P^{-1.18}$ (see inset Fig. 4).  The exact value of the exponent is
quite influenced by the error bar on $V_0$ since this latter
parameter is not very precise and is found to change very quickly
near $t=0$.

The temporisation stage can be attributed to the increase of
temperature in the system.  Indeed, a thermocouple has measured that
the inner temperature of the heap can increase by about 20 K,
depending on the injected current value.  This effect is
particularily noticeable for high currents, since more power has to
be dissipated.  The power dissipated (calculated at $t=$2200 s) by
the beads has been found to be 0.3, 1.9, 3.4 and 5.2 W for $i=$10,
25, 35 and 50 mA respectively.  Note that the local change in
temperature not only modifies the nature of contacts, but also the
structure of the network because of dilatation processes that may
lead to a reorganization of system structure.  This may cause to a
further increase in the resistance since microsolderings may be
damaged.

As seen before, the first decay seems to essentially concern the
enhancement of the conductance due to the growth of the conducting
network.  On the other hand, the long term decay seems to be related
to the enhancement of the contacts themselves.  The thin insulating
layer deteriorate with leakage current \cite{mich}.  Microsolderings
occur and the global resistance decreases.  The system can be easily
reset to its initial resistance by a tap on the vessel.

\section{Conclusions}

The above evidences a diffusion dynamic has been evidenced.  When a
current is injected through granular packing, it takes much time
(more than 2000 s) for the system to stabilize its resistance.  First
the resistance decreases due to a network effect : the well connected
cluster grows.  Thus, because of the power dissipation, the
temperature change induces dilatation of certain beads : the
resistance may increase according to the value of the power.
Finally, contacts are enhanced by the slow deterioration of the oxide
layer of beads.

To summarize, the behavior of such material is thus completely
different from a metallic bulk material.  The perfect example is what
happens in a lamp.  The lamp filament has a resistance of about 50
$\Omega$ for low current and at room temperature.  When the current
is switched on, the lamp produces light and heat.  This change in
temperature makes the electrical resistance increase.  On the other
hand, in the case of granular material, the opposite effect occurs.
When the current is first set, the resistance is high.  The heat, the
aging and some Zener effects affect the resistance by decreasing it.
Those effects contribute to the growing of the connected cluster of
beads.

Practically, as soon as some material presents weak links, electrical
measurements have to be carefully achieved.  The injected current
history and the injecting rate of the current are both seen to be
very relevant parameters.  Further theoretical works will be
developed on the basis of those observations in order to model such
complex electrical networks.

\section{Acknowledgements} S.D. would like to thank FNRS for
financial support.  This work has been also supported by the contract
ARC 02/07-293.  We would like to thank Prof. H.W. Vanderschueren for
the use of MIEL
facilities. We want also to thank Dr. A. B\'essinck (U. of Li\`ege) for his
valuable comments.  Fruitful discussions with Prof. G. Albinet and Dr. L.
Raymond are gratefully acknowledged.

\vskip 1cm \newpage \parindent=0pt \newpage \parindent=0pt

{\large \bf Figure Captions}

\vskip 0.5cm{\bf Figure 1}  Sketch of the $i-V$ behavior of a metallic granular
material

\vskip 0.5cm{\bf Figure 2} Evolution of the electrical resistance of a lead
beads heap with respect to time.  The different curves have been
obtained for several fixed injected currents as indicated by the
legend.  The symbols are placed every 50 points

\vskip 0.5cm{\bf Figure 3} Normalized
voltage versus time obtained for different currents (see legend).
One point out of 50 is indicated by a symbol; solid lines join the
data points

\vskip 0.5cm{\bf Figure 4} Normalized voltage $V/V_0$ versus time.  The
symbols corresponds to different injected currents (see legend) and
take place one data point over 10.  (inset) Relation between the
relaxation time $\tau$ and the injected current

\newpage
\begin{figure} \begin{center}
\centerline{\epsfxsize=8.6cm \epsffile{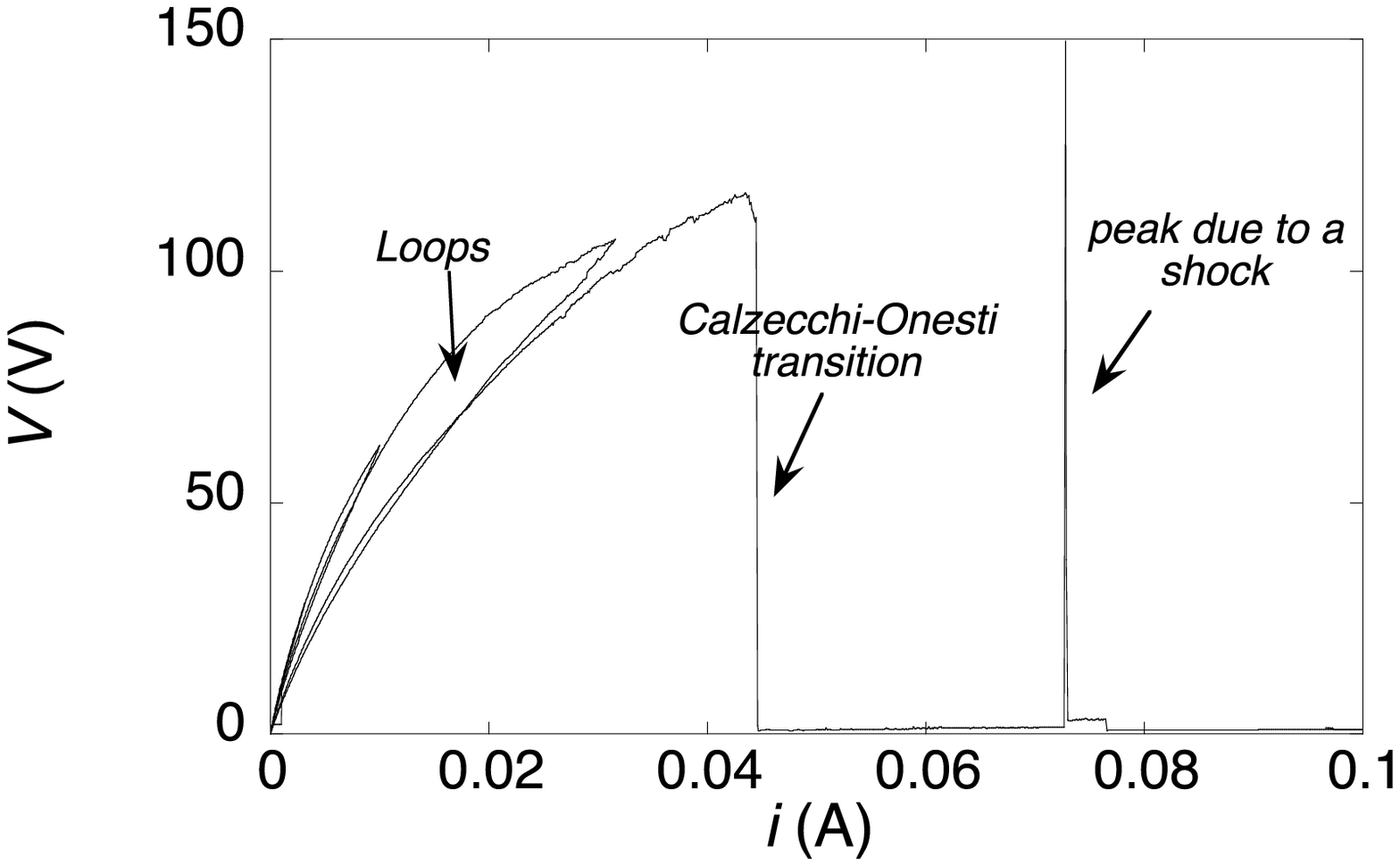}}
\caption{Dorbolo-JR03-1539}
\end{center} \end{figure}
\newpage
\begin{figure} \begin{center}
\centerline{\epsfxsize=8.6cm \epsffile{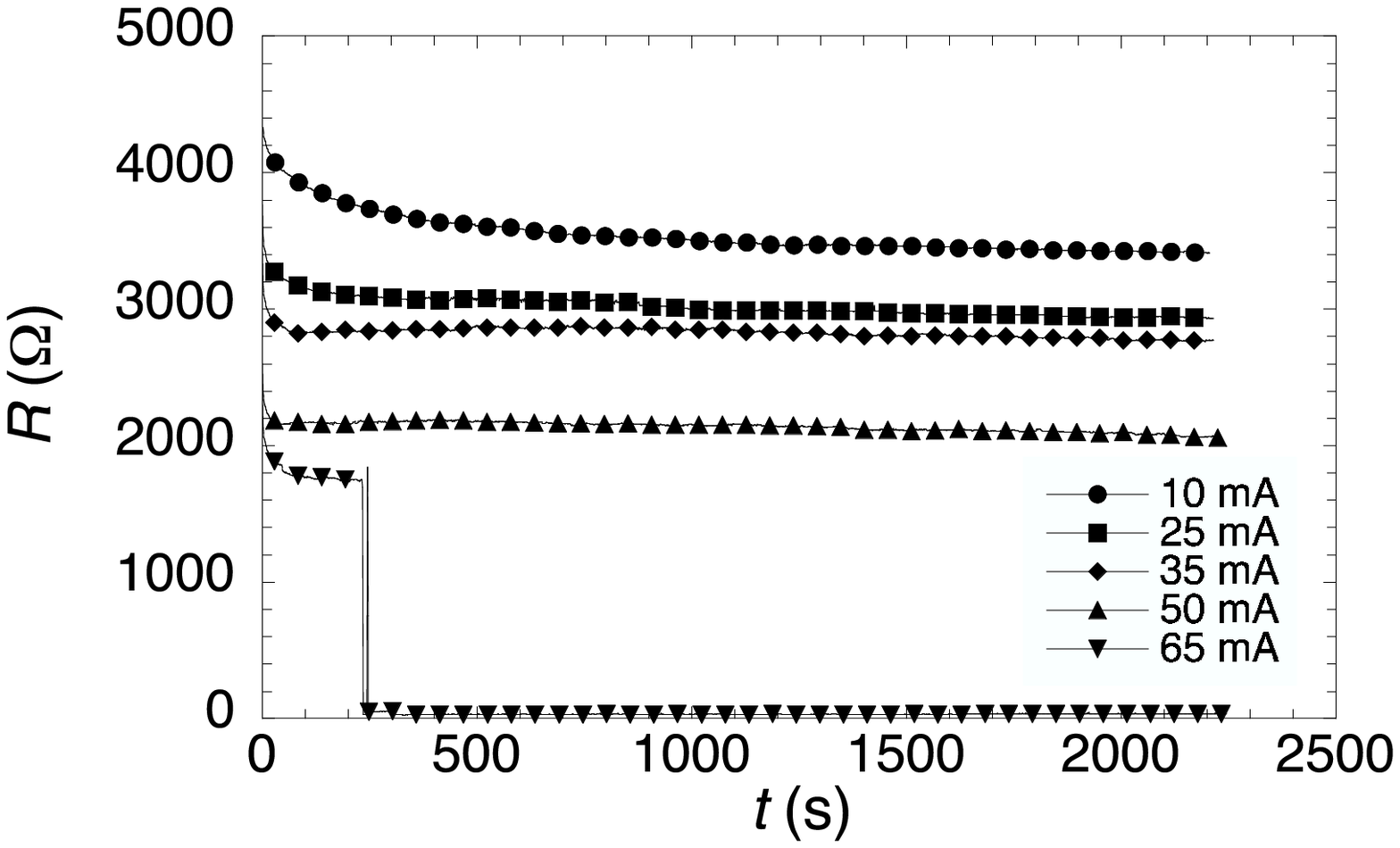}} \caption{Dorbolo-JR03-1539}
  \end{center} \end{figure}
\newpage
\begin{figure} \begin{center}
\centerline{\epsfxsize=8.6cm \epsffile{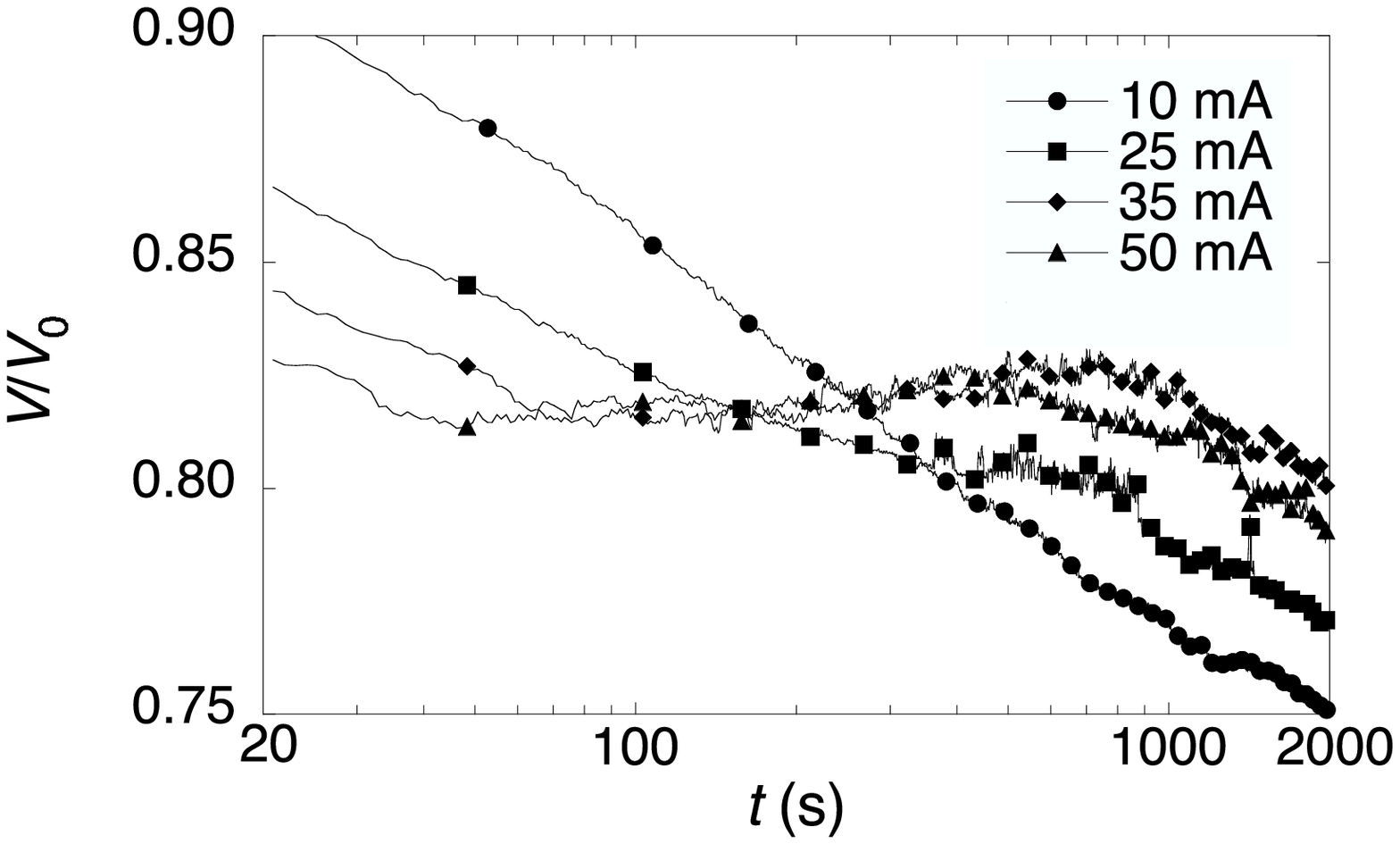}}
\caption{Dorbolo-JR03-1539}
\end{center} \end{figure}
\newpage
\begin{figure} \begin{center}
\centerline{\epsfxsize=8.6cm \epsffile{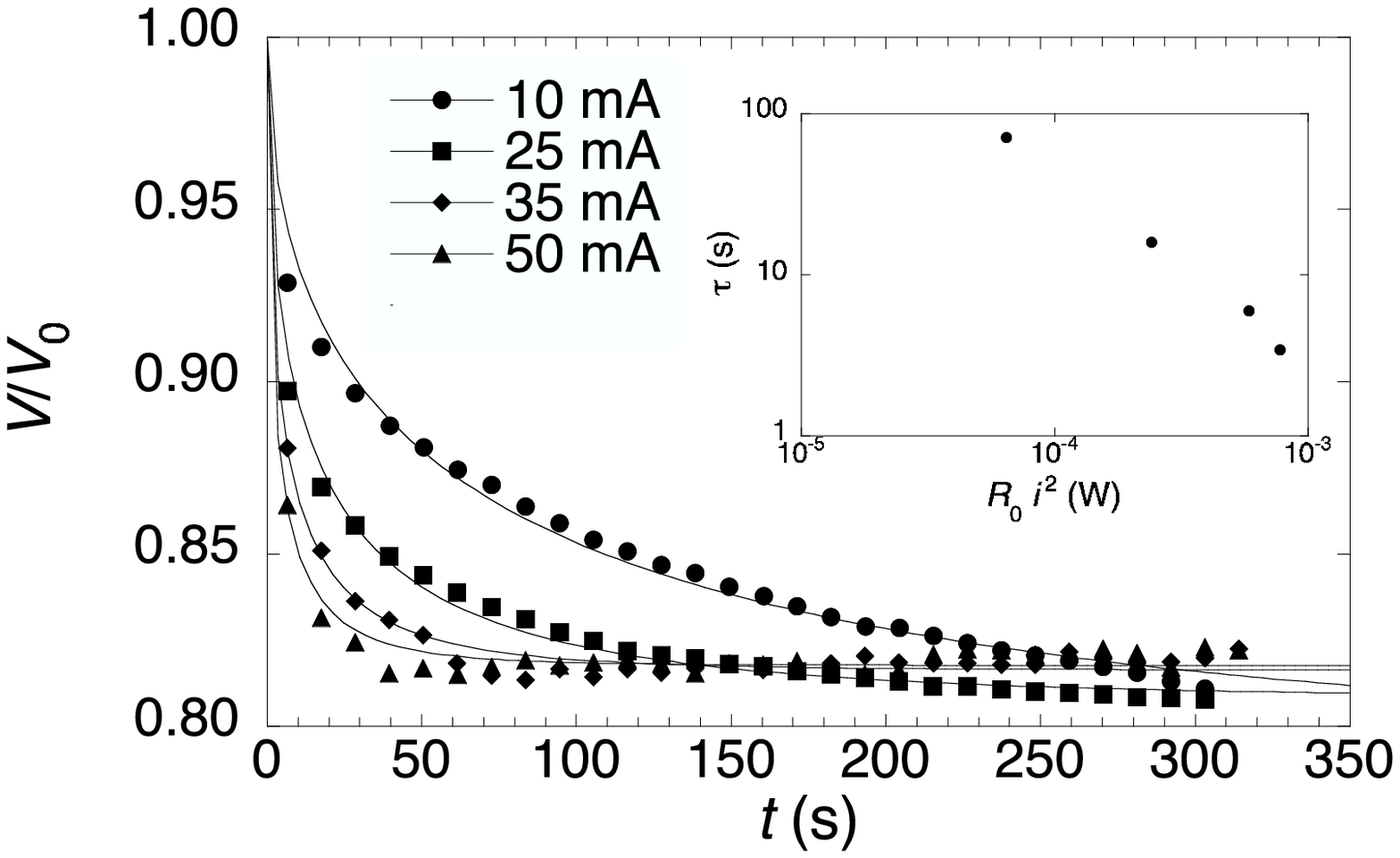}}
\caption{Dorbolo-JR03-1539}
\end{center} \end{figure}

\end{document}